\documentclass[journal]{vgtc}                     % final (journal style)
\usepackage{times}                     % we use Times as the main font
\usepackage{tabu}                      % only used for the table example
\usepackage{booktabs}                  % only used for the table example
\usepackage{lipsum}                    % used to generate placeholder text
\usepackage{mwe}                       % used to generate placeholder figures
\usepackage{mathptmx}                  % use matching math font
\usepackage{makecell}
\usepackage{multirow}
\usepackage{hhline}
\usepackage{graphicx}

\newif\ifshowrevs
\showrevstrue  
\ifshowrevs
  \newcommand{\rev}[1]{\textcolor{black}{#1}}
\else
  \newcommand{\rev}[1]{#1}
\fi

\onlineid{0}

\vgtccategory{Research}

\vgtcpapertype{please specify}

\title{Handows: A Palm-Based Interactive Multi-Window Management System in Virtual Reality}

\author{%
  \authororcid{Jin-Du Wang}{0009-0009-4028-4662},
  \authororcid{Ke Zhou}{0009-0007-9931-1616}, 
  \authororcid{Haoyu Ren}{0009-0004-5954-1167}, 
  \authororcid{Per Ola Kristensson}{0000-0002-7139-871X}, and
  \authororcid{Xiang Li}{0000-0001-5529-071X}
}

\authorfooter{
  \item
  	Jin-Du Wang is with The Hong Kong University of Science and Technology.
  	E-mail: jwangki@connect.ust.hk.
  \item
  	Ke Zhou is with Chongqing University.
  	E-mail: kezhou@stu.cqu.edu.cn.

  \item 
        Haoyu Ren is with Xi'an Jiaotong University.
  	E-mail: 2206113950@stu.xjtu.edu.cn.

  \item Per Ola Kristensson and Xiang Li are with the University of Cambridge.
  	E-mail: E-mail: \{pok21, xl529\}@cam.ac.uk.
  \item Xiang Li is the corresponding author.
}

\abstract{%
Window management in virtual reality (VR) remains a challenging task due to the spatial complexity and physical demands of current interaction methods. We introduce \textit{Handows}, a palm-based interface that enables direct manipulation of spatial windows through familiar smartphone-inspired gestures on the user's non-dominant hand. Combining ergonomic layout design with body-centric input and passive haptics, Handows supports four core operations: window selection, closure, positioning, and scaling. We evaluate Handows in a user study ($N=15$) against two common VR techniques (virtual hand and controller) across four core window operations. Results show that Handows significantly reduces physical effort and head movement while improving task efficiency and interaction precision. A follow-up case study ($N=8$) demonstrates Handows' usability in realistic multitasking scenarios, highlighting user-adapted workflows and spontaneous layout strategies. Our findings also suggest the potential of embedding mobile-inspired metaphors into proprioceptive body-centric interfaces to support low-effort and spatially coherent interaction in VR.
}


\teaser{
  \centering
 \includegraphics[width=\linewidth]{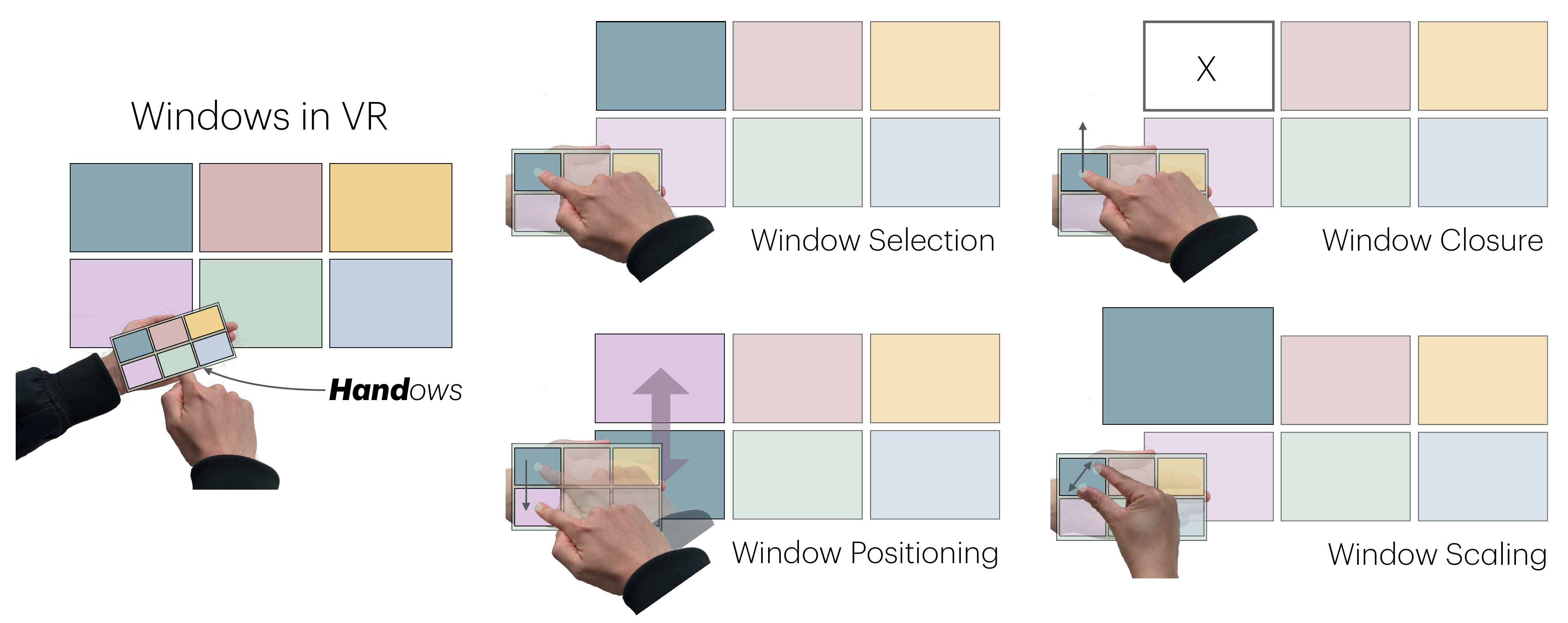}
  \caption{In this paper, we present \emph{Handows}---a palm-based interactive window management system designed for virtual reality (VR). It enables users to control multiple spatial windows in VR environments using their non-dominant palm as an interactive surface. The system supports four core operations: (1) window selection, (2) window closure, (3) window positioning, and (4) window scaling.}
  \label{fig:teaser}
}

\graphicspath{{figs/}{figures/}{pictures/}{images/}{./}} % where to search for the images

\usepackage{tabu}                      % only used for the table example
\usepackage{booktabs}                  % only used for the table example
\usepackage{lipsum}                    % used to generate placeholder text
\usepackage{mwe}                       % used to generate placeholder figures

\usepackage{mathptmx}                  % use matching math font

\begin{document}

%%%%%%%%%%%%%%%%%%%%%%%%%%%%%%%%%%%%%%%%%%%%%%%%%%%%%%%%%%%%%%%%
%%%%%%%%%%%%%%%%%%%%%% START OF THE PAPER %%%%%%%%%%%%%%%%%%%%%%
%%%%%%%%%%%%%%%%%%%%%%%%%%%%%%%%%%%%%%%%%%%%%%%%%%%%%%%%%%%%%%%%

\firstsection{Introduction}

\maketitle

One of the major promises of virtual reality (VR) environments is their ability to overcome physical constraints and offer virtually unlimited display space~\cite{monitor}. Unlike traditional monitors or mobile devices, VR systems allow users to open, position, and manipulate multiple application windows of varying sizes throughout a 3D virtual environment~\cite{eliot_1}. This spatial flexibility is especially appealing for multitasking and information-rich workflows. However, the current state of window management in VR is still limited in both usability and efficiency~\cite{seok_1}.

In traditional computing environments, such as desktops, laptops, and smartphones, window management is supported by mature 2D graphical user interfaces (GUIs), which rely on the well-established WIMP model: windows, icons, menus, and pointers~\cite{3DUI,WIMP}. These systems enable a wide range of functionalities, including window creation, movement, resizing, and layout customization~\cite{windowmanipulation}. Multi-window management is further supported by features like task overviews, layout preservation, and rapid switching between windows~\cite{TaskGallery,Xwindow,TaxonomyWindow}. Critically, the compatibility between these GUI elements and 2D input methods, such as mice or touchscreens, enables efficient interaction with minimal physical or cognitive effort~\cite{touchWindowManage}.

In contrast, VR platforms generally provide only basic interaction techniques for spatial windows~\cite{Biener_Breaking_20,Biener_Quantifying_22}, often relying on ray-casting with controllers or mid-air gestures for selection and manipulation~\cite{nealF_1,veronica_1,markR_1}. Although these methods offer spatial freedom, they often require large arm movements, precise coordination, and repetitive actions, which can lead to fatigue~\cite{eliot_1}. Moreover, the absence of tactile feedback and limited layout support hinders fluid multitasking and distracts users from their primary goals~\cite{markR_1}.

To address these challenges, we introduce \textit{Handows}, a novel window management system designed specifically for VR environments. Handows is a palm-based interface situated on the user's non-dominant hand, allowing direct interaction with window thumbnails through familiar gestures such as tapping, swiping, and pinch-to-zoom. Drawing inspiration from mobile interaction paradigms and grounded in principles of human-computer interaction (HCI)~\cite{mueller_towards_2023,floyd_mueller_limited_2021,Biener_PoVRPoint_22}, Handows aims to reduce physical strain and cognitive effort by integrating a miniature interface into the user's proprioceptive space. It also introduces a default ergonomic layout for organizing windows based on angular and distance-related comfort zones~\cite{seok_1,Grubert1}.

To evaluate the effectiveness of \textit{Handows}, we conducted a two-part investigation. First, a user study with 15 participants compared \textit{Handows} with a virtual hand interface and a controller-based baseline across four fundamental window operations: selection, closure, positioning, and scaling. Performance metrics included task completion time, accuracy, head and hand movement, and subjective preference. Second, we carried out a case study simulating a realistic multitasking scenario (i.e., trip planning and budgeting) to examine how users employ \textit{Handows} in continuous, goal-driven workflows.

Findings from both studies reveal that \textit{Handows} significantly improves task efficiency, reduces physical effort, and enhances user satisfaction. In the user study, Handows outperformed the other techniques in selection and closure tasks, reduced head rotation by over 60\%, and achieved the most precise scaling performance. In the case study, participants adapted familiar mobile-inspired interaction patterns and spontaneous layout strategies, reporting that the palm-based interface aligned closely with their interaction habits and supported fluid multitasking in immersive environments. In summary, this paper makes the following contributions:
\begin{itemize}
    \item We introduce \textit{Handows}, a palm-based VR window management system that integrates spatial miniaturization, passive haptics, and mobile-inspired gesture interaction \rev{within the user's proprioceptive space.}
    
    \item \rev{We demonstrate that embedding mobile-inspired metaphors into proprioceptive, body-centric interfaces improves user performance and reduces perceived effort in window management tasks in VR.}
    
    \item \rev{We identify how users adapt spatial layouts and develop sustained interaction strategies in a realistic multitasking scenario, revealing how body-anchored interfaces can support fluid, goal-driven workflows in immersive environments.}
\end{itemize}

\section{Related Work}

We situate Handows within three strands of prior research: (1) window management techniques in virtual environments, (2) spatial representation and miniature views, and (3) on-body interaction interfaces. Together, these threads inform the ergonomic, spatial, and interaction design of our system.

\subsection{Window Management in Virtual Environments}

In VR systems, application content is typically rendered as floating 2D windows within a 3D space~\cite{eliot_1}, which introduces new challenges for interaction and spatial organization~\cite{yipinghuang_1,seok_1}. Various input techniques have been explored. Direct hand-based interaction has been widely explored, but physical reach limitations and fatigue constrain its scalability~\cite{yipinghuang_1}. More commonly, ray-casting is used—either via hand-held controllers or gesture recognition~\cite{nealF_1,veronica_1,markR_1}—though these methods lack fine control and tactile grounding. To improve manipulation, Projective Windows~\cite{seok_1} introduced continuous hand gestures for window interaction, while cross-device AR systems have enabled basic window control via smartphones~\cite{yuanchunshi_1}. However, many of these approaches remain limited by their reliance on frequent arm movement and the absence of haptic cues.

Spatial layout also significantly impacts user performance and comfort. Egocentric layouts can reduce motion sickness and improve task efficiency~\cite{irani_1}, and users generally prefer multi-window configurations for complex tasks~\cite{bi_1}. While some systems aggregate windows into large immersive displays~\cite{leonardo_1}, others recommend curved or segmented arrangements to support peripheral awareness and multitasking~\cite{curve}. Recent work has also explored leveraging underutilized peripheral space and gaze-based cursor teleportation to improve window switching efficiency on large virtual displays~\cite{pavanatto2025spatial}, though such techniques still operate within conventional desktop interaction paradigms and offer limited ergonomic rethinking. Dynamic and context-aware window placements have also been proposed~\cite{fitzmaurice_1,david_1}, but such designs can increase the difficulty of spatial recall and require users to adapt to inconsistent spatial mappings~\cite{david_1}. 

\subsection{Miniature Representations and Spatial Navigation}

To support navigation and control in virtual spaces, several systems have introduced miniature representations. The World-in-Miniature (WIM) technique~\cite{stoakley_1} enables users to manipulate the environment at a reduced scale. Similarly, multi-viewport systems and minimap-style interfaces assist in orienting users to off-screen targets in both 2D and 3D~\cite{elvins_1,yu_1,tiago_1,burigat_1}. These representations improve global awareness but are often limited by visibility constraints or require additional workspace that may conflict with active content. While valuable for scene navigation, they are less frequently used for fine-grained content manipulation such as window control.

\subsection{On-Body Interaction and Passive Haptics}

On-body interaction offers a promising alternative to traditional mid-air gestures by leveraging proprioception and passive haptic feedback. Prior work has demonstrated that skin surfaces can serve as effective input regions, improving accuracy and reducing fatigue~\cite{fang_1,patibanda2023auto,patibanda2023fused,li2024onbodymenu,li2025Bend}. Systems like SkinWidget~\cite{azai_1} and others~\cite{harrison_1,harrison_2} implement forearm- and wrist-based input, although fatigue remains a concern with extended use. Similarly, microGEXT~\cite{li2025evaluating} leverages the side of the finger for sliding gestures, enabling self-haptic feedback during text selection and editing in VR.

Palm-based interaction, in particular, has shown strong potential due to its visual accessibility and biomechanical stability. Several projects have proposed projecting interactive elements onto the palm~\cite{wang_2,whitmire_1,chatain_1}, while others have explored imaginary touch interfaces~\cite{gustafson_1} or input techniques such as PalmGesture~\cite{wang_1} and on-body menus~\cite{azai_2,li2024onbodymenu}. These works suggest that the palm can act as a natural, proprioceptively-aligned interaction surface. STAR~\cite{STAR} demonstrates how smartphone-like interactions can be effectively mapped to the hand, pointing to opportunities for integrating familiar 2D paradigms into immersive contexts. 

In summary, our work builds on insights from VR window management, spatial navigation tools, and on-body input research. \rev{ While prior systems have explored the palm as an interactive surface  and used miniature representations for spatial navigation, the novelty of Handows lies in its specific synthesis of these concepts for the complex task of multi-window management in VR. Handows contributes a unique combination of spatial miniaturization, passive haptics from the user's own body, and familiar smartphone-inspired gestures, all situated within an ergonomically optimized layout. This integrated, body-centric approach is designed to provide more usable and efficient VR window control.}

\section{Handows: A Window Management System for VR}

\begin{figure}[h]
    \centering
    \includegraphics[width=\linewidth]{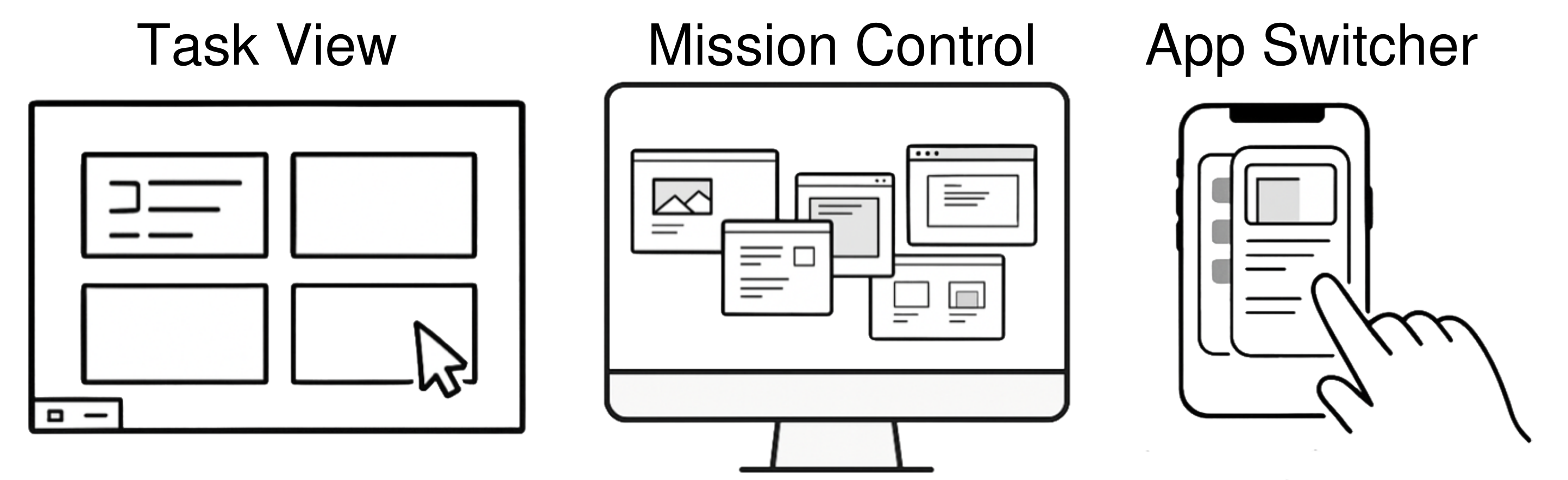}
    \caption{Common multi-window management paradigms across platforms. \textbf{Task View} (Windows) and \textbf{Mission Control} (macOS) provide spatial overviews of open windows using mouse input. On mobile platforms, the \textbf{App Switcher} allows users to navigate recent tasks through swipe gestures. These systems inspired the design of \textit{Handows}, which adapts structured layout and gesture-based control into a palm-based interface for VR.}
    \label{fig:inspiration}
\end{figure}

Modern operating systems provide robust and intuitive solutions for managing multiple application windows. On Windows PCs, \textit{Task View} enables users to overview and switch between tasks using a structured, tiled layout. macOS offers \textit{Mission Control}, which presents all open windows spatially across desktops, facilitating spatial reasoning and task switching. On mobile platforms, both Android and iOS implement \textit{App Switcher} views that support fluid navigation across applications through touch gestures like swipe and tap (see Figure~\ref{fig:inspiration}).

These existing paradigms emphasize spatial structure and rapid access that we aim to bring into immersive environments. Inspired by these systems, we introduce \textit{Handows}, a palm-based window management system designed specifically for VR. Handows addresses the spatial, ergonomic, and attentional challenges of managing multiple windows in 3D by anchoring the interface on the user's non-dominant palm. The system integrates spatial miniaturization, passive haptic feedback, and gesture-based control to support fast, embodied interaction with minimal fatigue. At its core, Handows consists of two key components: (1) a default spatial layout that helps users organize and position multiple windows efficiently in the environment, and (2) a palm-based miniature interface that supports core operations—window selection, closure, positioning, and scaling—through smartphone-inspired gestures.

\subsection{Spatial Layout Optimization}

While VR enables the placement of an unlimited number of application windows throughout 3D space, unrestricted spatial freedom often leads to disorganized layouts, increasing cognitive load and interaction complexity~\cite{contextaware}. To mitigate these issues, Handows adopts a default window layout designed to promote visual clarity, reduce head movement, and enhance access efficiency.

The system follows a multi-display strategy in which each application occupies an independent virtual window, rather than combining content into a unified display surface~\cite{bi_1,gao2023vr}. Drawing from ergonomic guidelines on monitor positioning~\cite{Microsoft2021Comfort,CCOHS2022MonitorPositioning}, Handows places windows within the user's natural field of view, constrained by angular and distance thresholds to reduce visual fatigue. The central window is reserved for the user's primary task, while secondary windows are placed at similar or higher elevations, avoiding vertical overlap with physical furniture such as desks~\cite{monitor}. This spatial configuration maintains consistency with real-world expectations while minimizing interference and supporting fluid transitions between windows.

\subsubsection{Curved Layout Selection}

\begin{figure}[t]
  \centering
  \includegraphics[width=\linewidth]{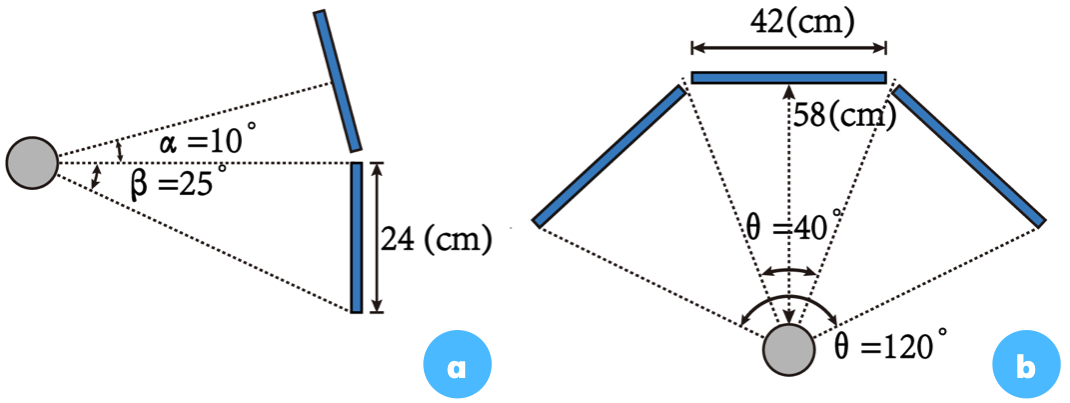}
  \caption{Representations of the default layout (a) Side view, (b) Top view. All screens are placed within an area where the user's head rotation does not exceed 45 degrees to the left or right and 10 degrees upward~\cite{Microsoft2021Comfort}. The primary screen is positioned in the optimal area~\cite{CCOHS2022MonitorPositioning,monitor}, and each screen is 19 inches (a common size of displays).}
  \label{fig:Description}
\end{figure}

To refine the spatial configuration, we conducted an in-lab design parameter study ($N=8$) comparing four alternative layout geometries: flat, horizontally curved, vertically curved, and both horizontally and vertically curved. Eight participants in a controlled lab setting evaluated each layout in terms of usability and peripheral search efficiency. Following Barrett et al.'s task design~\cite{personalcockpit}, participants performed target-counting tasks involving color-shape combinations across peripheral displays. Each participant completed 20 trials under each layout condition. Layout order was counterbalanced to mitigate ordering effects. Our results indicated a consensus preference for the combined horizontal and vertical curvature layout. Participants reported improved visibility and reduced neck rotation during target acquisition. Based on these findings, Handows adopts this curved configuration as the system's default spatial arrangement (see Figure~\ref{fig:Description}).

\subsection{Palm-Based Interface and Interaction Design}

Handows situates its interface on the user's non-dominant palm, emulating the size and feel of a smartphone screen. This body-centric approach offers both proprioceptive alignment and passive haptic feedback, enhancing spatial awareness and reducing visual dependence during interaction. The interface is activated when the palm faces upward and displays a grid of window thumbnails, each representing a currently open application. This layout supports simultaneous access to multiple windows without overwhelming the user's visual field.

\begin{figure}[t]
  \centering
  \includegraphics[width=\linewidth]{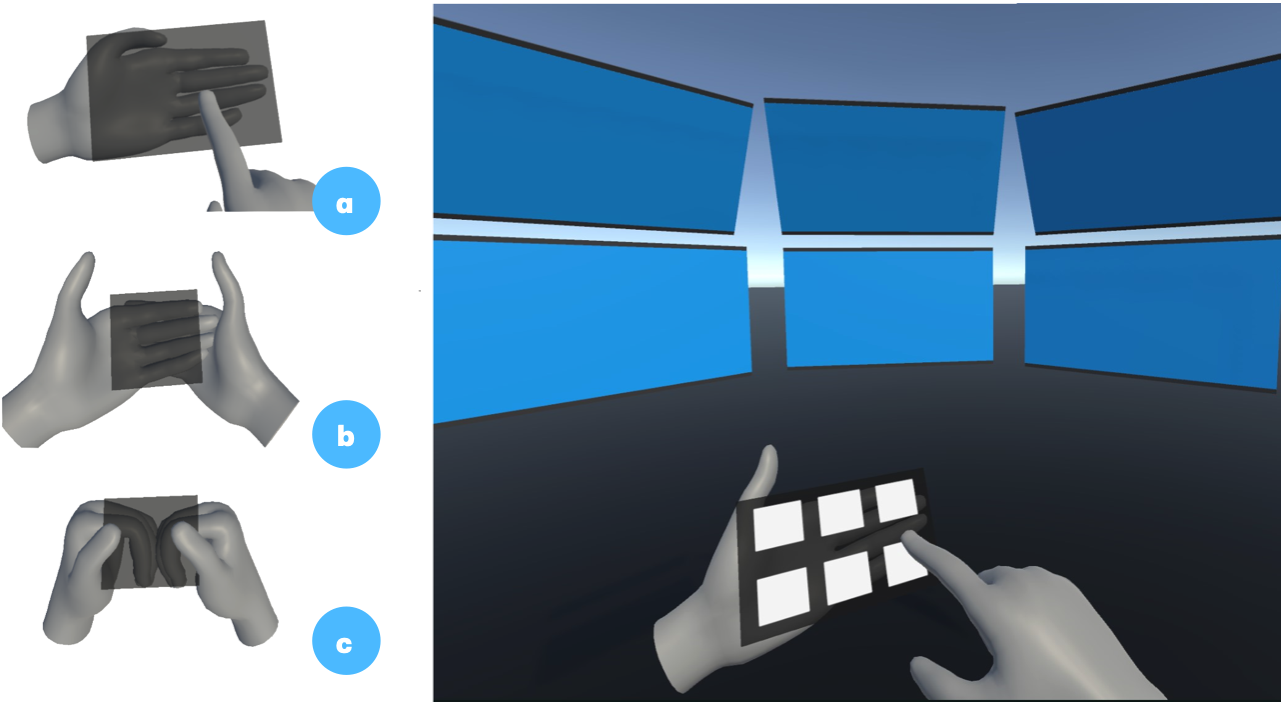}
  \caption{Left: Three surfaces providing passive haptic feedback with operational panels (black areas) for virtual reality interactions: (a) Single Palm Surface (opted in Handows), (b) Nested Finger Surface, (c) Index Finger Surface. Right: Overview of the Handows interface, illustrating the comprehensive layout and functional components.}
  \label{fig:DesignHandows}
\end{figure}

We considered alternative placements, including finger-based surfaces and nested configurations~\cite{STAR} (see Figure~\ref{fig:DesignHandows} (b) and (c)), but selected the palm interface following exploratory testing, which indicated advantages in interaction area, posture stability, and visual accessibility. The interface encourages a natural phone-holding posture, facilitating prolonged use while minimizing fatigue.

\subsection{Features}

Handows supports four core window operations: (1) selection; (2) closure; (3) positioning; and (4) scaling (see Figure~\ref{fig:teaser}), by adapting familiar mobile gestures to embodied interaction in VR. All interactions are performed using the dominant hand, while the non-dominant palm serves as a stable, proprioceptively accessible control surface. This bimanual configuration minimizes fatigue and supports fluid transitions between operations.

\begin{itemize}
    \item \textbf{Selection:} Users tap a window thumbnail on the palm interface to bring the corresponding window into focus.
    \item \textbf{Closure:} A swiping gesture toward the edge of the palm removes the selected window from the layout.
    \item \textbf{Positioning:} Dragging a thumbnail repositions the associated window, which automatically snaps to the \rev{nearest predefined slot} in the ideal spatial layout.
    \item \textbf{Scaling:} Pinch-to-zoom gestures are used to resize windows, mimicking zoom interactions on smartphones.
\end{itemize}

By integrating well-established gesture paradigms within a body-centric interface, Handows enables intuitive and efficient window management in immersive environments. This combination of mobile familiarity, passive haptic feedback, and ergonomic layout design supports low-effort, high-control interaction across a range of VR workflows. \rev{To ensure a rigorous evaluation, the system's scope is deliberately focused on the manipulation of existing windows. This choice avoids the complex and confounding variables tied to new window creation. Supporting the latter would require addressing broader issues, such as where new windows should appear, how to resolve overlaps, and how to group or order windows, which we consider beyond the scope of this work.}

\section{User Study}
To assess the performance and user experience of Handows for window management tasks in VR environments, we carried out a user study comparing it with two baseline techniques prevalent in current VR systems: Virtual Hand and Controller. This investigation aimed to examine how the design characteristics of Handows, such as body-centric anchoring, spatial miniaturization, and gesture familiarity, affect user interaction compared to established methods. Our goal was to evaluate not only interaction efficiency, but also physical demand and subjective experience. Although the Handows design is applicable to both VR and MR environments, we conducted our experiments in VR to ensure environmental control and consistency, thereby strengthening the internal validity of our findings. \rev{To guide this study, we proposed the following hypotheses:}

\rev{\textbf{\textit{H1}}: We hypothesized that by adapting familiar mobile gestures onto a miniaturized, palm-based interface, \textsc{Handows} would enable faster and more precise window manipulation compared to existing techniques.}

\rev{\textbf{\textit{H2}}: We expected that anchoring the interface to the user's body would significantly reduce physical efforts and fatigue during window management tasks.}

\rev{\textbf{\textit{H3}}: We anticipated that the combination of passive haptic feedback, proprioceptive stability, and the high transferability of familiar gestures would make \textsc{Handows} a more intuitive and satisfying technique, resulting in more positive subjective experiences and stronger user preference for \textsc{Handows} over the baselines.}

\subsection{Participants and Apparatus}
A total of 15 participants (10 males and 5 females) were recruited for the study. The age range of the participants was between 20 and 23 years (\(M = 20.67, SD = .72\)). All participants were students at a local university. All participants reported previous experience with VR, with familiarity ratings ranging from 1 to 7 on a 7-point Likert scale, where 1 indicated no experience in VR, and 7 indicated expertise (\(M = 3.13, SD = 1.96\)). All participants were right-handed habitual users. This study was approved by the XJTU research ethics board.

The user study was conducted in a university laboratory equipped with a desktop computer, display devices, and an area for participants to engage in VR interactions. The application, developed in Unity 2022.3.8f1, was run from the Unity Editor on a desktop PC and streamed to a Meta Quest 2 headset \rev{via a wired Quest Link connection.} The system utilized the Meta Interaction SDK for hand tracking. \rev{Window management operations recognition was based on detecting a collision between the fingertip of the dominant hand and a virtual panel on the non-dominant hand. This method is robust to variations in resting hand pose (e.g., whether the hand is clenched or extended) and does not rely on a specific finger configuration. Furthermore, the Meta Interaction SDK automatically estimates the user's hand size during initialization and scales the virtual hand model accordingly.}

\subsection{Method and Procedure}

\begin{figure}[t]
  \centering
  \includegraphics[width=\linewidth]{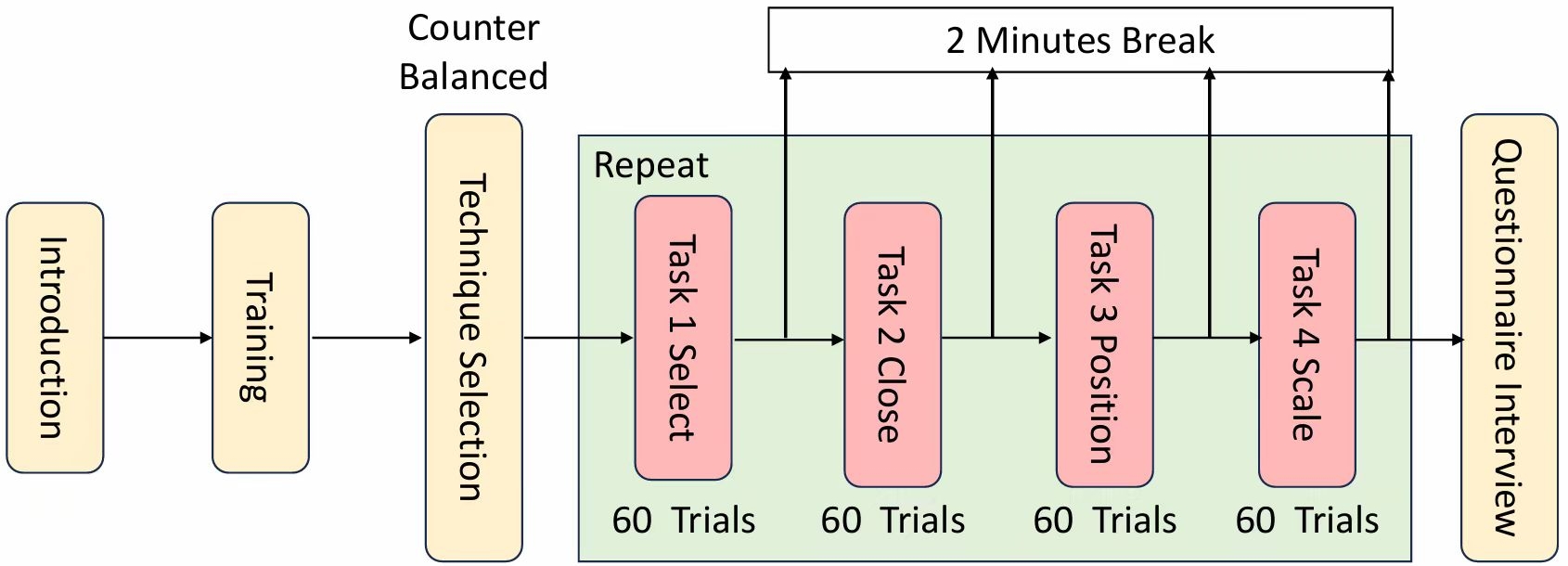}
  \caption{The procedure of the user study.}
  \label{fig:StudyProcess}
\end{figure}

The study followed a within-subject design with one independent variable, \textsc{window management technique}. Users were tasked with completing four objectives: target window selecting, closing, positioning, and scaling, using three different window management techniques: Handows, Virtual Hand, and Controller.

After a brief introduction, participants were given 10 minutes to familiarize themselves with the window management techniques and tasks that would be undertaken. Additionally, based on each user's dominant hand and operating habits, it was confirmed which hand would serve as the operating hand, while the other hand would be used for placing the control panel.

To minimize the influence of learning effects, the order of window management techniques assigned to participants was counterbalanced using a Latin square design. After participants completed the designated tasks for each technique, they were given a two-minute break before proceeding to the next technique. After completing all tasks, participants were asked to fill out questionnaires evaluating their experience with the different manipulation techniques. Further, a final comparison was included, where participants were asked to rate their preference for each manipulation technique. Finally, a semi-structured interview was conducted to gather qualitative insights, allowing participants to share their strategies and provide suggestions. The process is shown in Figure \ref{fig:StudyProcess}. On average, participants spent approximately 50 minutes completing the study, including the tasks and questionnaire. Participants received a small monetary reward for their participation.

\subsection{Tasks and Measurement}
The evaluation metrics include: (a) \textit{Task Completion Time} (s), measuring the duration from the target window's appearance to task completion; (b) \textit{Hand Movement Distance} (m), representing the total distance traveled by \rev{the participant's dominant hand}; (c) \textit{Head Rotation Angle} (°), indicating the degree of head movement; and specifically for Task~4, (d) \textit{Scaling Deviation} (\%), which captures the difference between the actual and target window sizes to assess scaling accuracy, as suggested by Li et al.~\cite{li2023swarm,li2024swarm}.

We also collected subjective feedback through several questionnaires: (a) \textit{Perceived Workload}, measured by the raw NASA-TLX~\cite{hart_nasa-task_2006}; (b) \textit{Usability}, measured by the System Usability Scale (SUS)~\cite{Brooke}; (c) \textit{User Experience}, measured by the UEQ-Short~\cite{schrepp2017ueq}; (d) \textit{Perceived Fatigue}, measured by the Borg 6–20 scale~\cite{borg}; and (e) \textit{Simulator Sickness}, measured by the Simulator Sickness Questionnaire (SSQ)~\cite{Kennedy1993SimulatorSQ}. Finally, participants were asked to provide a comparative evaluation of the three techniques through a final questionnaire. This included: (1) ranking their preferred window management technique among Handows, Virtual Hand, and Controller; (2) rating the impact of passive haptic feedback from touching the hand during Handows operation on operational stability (on a 7-point scale); and (3) assessing the extent to which prior experience with traditional computing devices, such as computers and smartphones, could be transferred to each technique (on a 7-point scale).

\subsubsection{Task 1: Target Window Selection}
In this task, participants were asked to select a target window, which was highlighted in red at the start of each trial. Upon selection, the window reverts to its original color, and a new target is highlighted. Participants were instructed to perform ten selection operations for each of the six windows, totaling 180 selections per participant across three conditions. The appearance order of the target windows was randomized. 

\begin{table*}[t]
\centering
\caption{Mean values (M), standard deviations (SD), and test statistics for completion time, hand movement distance, and head rotation across the four tasks (Target Selection, Closure, Positioning, and Scaling) using three interaction techniques: \textsc{Handows}, \textsc{Controller}, and \textsc{Virtual Hand}. For each metric, either repeated-measures ANOVA ($F$) or Friedman test ($\chi^2$) was conducted based on the normality of data distribution. Significant $p$-values are reported for overall effects; post-hoc test results are detailed in the text.}
\label{tab:result}
\resizebox{\textwidth}{!}{
\begin{tabular}{@{}llcccccc@{}}
\toprule
\multirow{2}{*}{Task} & \multirow{2}{*}{Technique} & \multicolumn{2}{c}{Time} & \multicolumn{2}{c}{Hand Movement Distance} & \multicolumn{2}{c}{Head Rotation Angle} \\
\cmidrule(lr){3-4} \cmidrule(lr){5-6} \cmidrule(l){7-8}
& & M (SD) & Statistics & M (SD) & Statistics & M (SD) & Statistics \\
%& & M (SD) & Stat. & M (SD) & Stat. & M (SD) & Stat. \\
\midrule
\multirow{3}{*}{Task 1} 
& Handows       & 0.88 (0.16) & $p < 0.0001$ & 0.22 (0.07) & $p < 0.0001$ & 4.67 (1.07)  & $p < 0.0001$ \\
& Controller    & 0.92 (0.15) & $F(2,28) = 28.271$ & 0.21 (0.04) & $\chi^2(2) = 19.6$ & 23.46 (8.11) & $F(2,28) = 58.831$ \\
& Virtual Hand  & 1.18 (0.23) & & 0.34 (0.09) & & 29.78 (10.35) & \\
\midrule
\multirow{3}{*}{Task 2} 
& Handows       & 1.16 (0.24) & $p < 0.0001$ & 0.28 (0.10) & $p < 0.0001$ & 6.42 (2.98)  & $p < 0.0001$ \\
& Controller    & 1.50 (0.34) & $F(2,28) = 96.413$ & 0.24 (0.07) & $F(2,28) = 142.498$ & 28.38 (7.78) & $\chi^2(2) = 26.53$ \\
& Virtual Hand  & 2.53 (0.58) & & 0.36 (0.07) & & 37.20 (6.82) & \\
\midrule
\multirow{3}{*}{Task 3} 
& Handows       & 1.88 (0.44) & $p < 0.0001$ & 0.30 (0.12) & $p < 0.0001$ & 7.38 (3.09)  & $p < 0.0001$ \\
& Controller    & 1.30 (0.21) & $F(2,28) = 28.695$ & 0.43 (0.12) & $\chi^2(2) = 24.13$ & 34.49 (18.55) & $F(2,28) = 54.255$ \\
& Virtual Hand  & 1.97 (0.27) & & 0.66 (0.15) & & 46.24 (12.14) & \\
\midrule
\multirow{3}{*}{Task 4} 
& Handows       & 2.10 (0.57) & $p < 0.0001$ & 0.20 (0.08) & $p < 0.0001$ & 7.76 (3.70)  & $p < 0.0001$ \\
& Controller    & 2.83 (0.57) & $F(2,28) = 110.622$ & 0.31 (0.05) & $F(2,28) = 56.522$ & 40.87 (7.50) & $\chi^2(2) = 28.13$ \\
& Virtual Hand  & 5.06 (1.13) & & 0.47 (0.09) & & 49.41 (8.38) & \\
\bottomrule
\end{tabular}
}
\end{table*}

\subsubsection{Task 2: Target Window Closure}
In this task, participants were asked to close the target window, which was highlighted in red. Upon closure, a new target window would appear among the remaining windows and would be highlighted in red. When all windows in the scene are closed, the scene resets and a new target window appears among the newly generated six windows. In each condition, participants were instructed to perform 10 closing operations on each of the six windows in the space, totaling 180 closing operations per participant. The appearance order of the target windows was randomized.

\subsubsection{Task 3: Target Window Positioning}
In this task, participants were asked to position a target window highlighted in red to a specified position highlighted in yellow. Upon successful positioning, the colors reverted, and new target windows and positions would be highlighted in red and yellow, respectively. In each condition, users were instructed to position each of the six windows in the space to the rest five positions (excluding the original position) twice, totaling 180 movement operations per user. The appearance order of the target windows and positions was randomized.

\subsubsection{Task 4: Target Window Scaling}
In this task, participants were required to scale the target window highlighted in red with a dashed outline indicating the desired size. If the current size was within a 5\% error range of the target size when the user completes the scaling operation, the current trial was considered completed. The original window size would be restored, and a new target window and target size would be generated. Each participant performed five scaling operations in both zoom-in (1.2 $\times$ original size) and zoom-out (0.8 $\times$ original size) modes on each window \rev{as suggested by Li et al.~\cite{li2023swarm,li2024swarm}}, totaling 180 operations per participant. \rev{The scaling factors were chosen to provide adequate fingertip travel distance for the palm-based gesture, while simultaneously avoiding significant visual clutter and inter-window occlusion.} The appearance orders of the target windows and sizes were randomized.

\subsection{Window Management Techniques}
In this study, participants were assigned assessment tasks utilizing three distinct techniques. Notably, Virtual Hand and Controller served as baseline methodologies, aligning with the default window management interactions of the Meta Horizon OS~\cite{li2023swarm,li2024swarm}. \rev{A key methodological adjustment was made for a fair comparison. Unlike the standard interactions of Meta Horizon OS, which restricts manipulation to window borders to avoid conflicting with in-window content interactions, our baselines permitted interaction with the entire window surface. This decision equalized target acquisition difficulty with the conflict-free Handows technique. The functional outcomes of all window operations were standardized across all three conditions.} The operational procedures for each baseline technique are delineated as follows:

\subsubsection{Virtual Hand (Ray-casting and Gesture)}
For the Virtual Hand technique, participants use a pinch gesture when the ray projected from the center of the palm intersects with the target window area to select it. Releasing the pinch confirms the selection. Similarly, closing a target window involves a pinch gesture when the ray aligns with the close button, and releasing the pinch confirms closure. To move a target window, participants utilize a pinch gesture when the ray intersects with the window area and then the selected window follows the hand's movement. Releasing the pinch halts the window's movement. Participants use a pinch gesture within the window's border area to resize a target window. Dragging outward enlarges the window while dragging inward reduces its size.

\subsubsection{Controller (Ray-casting and Button)}
For the Controller technique, participants interact with windows by positioning the ray projected from the front end of the controller onto the target window area to select it. Selection is initiated by pressing the trigger button, with confirmation upon trigger release. Similarly, closing a target window is executed by positioning the ray onto the close button and pressing the trigger button, confirming closure upon trigger release. To move a target window, participants position the ray onto the window area and hold the trigger button to enable the window to track the controller's movement. Releasing the trigger button halts the window's movement. Resizing operations are initiated by positioning the ray onto the window border and pressing the trigger button. Participants then drag outward to enlarge the window or inward to reduce its size, confirming the action upon trigger release.

\subsection{Results}

We analyzed both objective and subjective measures across all tasks and conditions. Normality was assessed using the Shapiro–Wilk test. For normally distributed variables, we applied repeated-measures ANOVA (RM-ANOVA) with Tukey HSD post-hoc comparisons. For non-normally distributed variables, we used the Friedman test with pairwise Mann–Whitney $U$ tests and Bonferroni correction. The significance level was set at $p < .05$. Descriptive statistics are summarized in Table~\ref{tab:result}.

\subsubsection{Task 1: Target Window Selection}

\paragraph{Task Completion Time.}
A significant main effect was found ($F(2, 28) = 28.27$, $p < .001$). Post-hoc tests showed that both \textsc{Handows} and \textsc{Controller} were significantly faster than \textsc{Virtual Hand} ($p = .0008$, $p = .0001$), with no difference between the former two ($p = .804$).

\paragraph{Hand Movement Distance.}
Friedman test revealed a significant effect ($\chi^2(2) = 19.60$, $p < .001$). Post-hoc results indicated that \textsc{Handows} and \textsc{Controller} required shorter distances than \textsc{Virtual Hand} ($p = .0008$, $p < .0001$). No difference was found between \textsc{Handows} and \textsc{Controller} ($p = .320$).

\paragraph{Head Rotation Angle.}
A significant difference was observed ($F(2, 28) = 58.83$, $p < .001$). \textsc{Handows} resulted in significantly smaller head rotation than both \textsc{Controller} and \textsc{Virtual Hand} (both $p < .0001$). The difference between \textsc{Controller} and \textsc{Virtual Hand} was not significant ($p = .071$).

\subsubsection{Task 2: Target Window Closure}

\paragraph{Task Completion Time.}
A main effect was found ($F(2, 28) = 96.41$, $p < .001$). \textsc{Handows} was significantly faster than both \textsc{Controller} ($p = .009$) and \textsc{Virtual Hand} ($p < .0001$). \textsc{Controller} also outperformed \textsc{Virtual Hand} ($p < .0001$).

\paragraph{Hand Movement Distance.}
Significant differences were found ($F(2, 28) = 142.50$, $p < .001$). \textsc{Handows} and \textsc{Controller} required less hand movement than \textsc{Virtual Hand} ($p = .026$, $p = .0003$).

\paragraph{Head Rotation Angle.}
Friedman test revealed a significant effect ($\chi^2(2) = 26.53$, $p < .001$). \textsc{Handows} elicited significantly smaller head rotation than both \textsc{Controller} and \textsc{Virtual Hand} (both $p < .0001$). \textsc{Controller} also differed from \textsc{Virtual Hand} ($p = .0004$).

\subsubsection{Task 3: Target Window Positioning}

\paragraph{Task Completion Time.}
A significant main effect was found ($F(2, 28) = 28.70$, $p < .001$). \textsc{Controller} was faster than both \textsc{Handows} and \textsc{Virtual Hand} (both $p < .0001$). No difference was observed between \textsc{Handows} and \textsc{Virtual Hand} ($p = .732$).

\paragraph{Hand Movement Distance.}
The effect was significant ($\chi^2(2) = 24.13$, $p < .001$). \textsc{Handows} outperformed both \textsc{Controller} ($p = .002$) and \textsc{Virtual Hand} ($p < .0001$). \textsc{Controller} also outperformed \textsc{Virtual Hand} ($p = .0002$).

\paragraph{Head Rotation Angle.}
ANOVA revealed a significant effect ($F(2, 28) = 54.26$, $p < .001$). \textsc{Handows} involved less head rotation than both \textsc{Controller} and \textsc{Virtual Hand} (both $p < .0001$), and \textsc{Controller} also differed from \textsc{Virtual Hand} ($p = .043$).

\subsubsection{Task 4: Target Window Scaling}

\paragraph{Task Completion Time.}
A significant effect was found ($F(2, 28) = 110.62$, $p < .001$). \textsc{Handows} was significantly faster than both \textsc{Controller} ($p = .031$) and \textsc{Virtual Hand} ($p < .0001$).

\paragraph{Hand Movement Distance.}
Differences were significant ($F(2, 28) = 56.52$, $p < .001$). \textsc{Handows} required less hand movement than both \textsc{Controller} ($p = .0008$) and \textsc{Virtual Hand} ($p < .0001$).

\paragraph{Head Rotation Angle.}
Friedman test showed a significant effect ($\chi^2(2) = 28.13$, $p < .001$). \textsc{Handows} resulted in significantly smaller head rotation compared to both \textsc{Controller} ($p = .0002$) and \textsc{Virtual Hand} ($p < .0001$).

\paragraph{Scaling Deviation.}
In Task 4, we additionally measured scaling deviation. A repeated-measures ANOVA revealed a significant main effect of interaction technique on deviation ($F(2, 28) = 41.60$, $p < .001$). \textsc{Handows} ($M = 1.39\%$, $SD = 0.22\%$) resulted in significantly more accurate scaling compared to both \textsc{Controller} ($M = 2.02\%$, $SD = 0.25\%$) and \textsc{Virtual Hand} ($M = 2.04\%$, $SD = 0.24\%$), with both comparisons reaching significance ($p < .0001$).

\subsubsection{Questionnaire Results}

\paragraph{User Experience (UEQ-S).}
Significant differences were found in Pragmatic Quality ($\chi^2(2) = 17.16$, $p = .0002$), Hedonic Quality ($\chi^2(2) = 10.56$, $p = .005$), and Overall Score ($\chi^2(2) = 14.88$, $p = .0006$). Post-hoc comparisons showed that \textsc{Handows} and \textsc{Controller} significantly outperformed \textsc{Virtual Hand} in Pragmatic Quality ($p = .001$, $p = .0004$, respectively). \textsc{Handows} also exceeded both alternatives in Hedonic Quality ($p < .0006$ vs. \textsc{Controller}, $p < .002$ vs. \textsc{Virtual Hand}). For Overall Score, only \textsc{Handows} significantly outperformed \textsc{Virtual Hand} ($p = .001$).

\paragraph{Perceived Workload (NASA-TLX).}
Significant differences were observed in Mental Demand, Physical Demand, Performance, Effort, and Frustration ($p < .01$ for all), but not in Temporal Demand ($p = .216$). \textsc{Virtual Hand} consistently induced higher workload across all significant dimensions when compared to both \textsc{Handows} and \textsc{Controller} ($p < .05$). Notably, \textsc{Handows} also required significantly less effort than \textsc{Controller} ($p = .003$).

\paragraph{Perceived Fatigue (Borg Scale).}
Fatigue scores differed significantly across techniques ($\chi^2(2) = 16.93$, $p = .0002$). Participants reported significantly higher fatigue with \textsc{Virtual Hand} compared to both \textsc{Handows} ($p < .0001$) and \textsc{Controller} ($p = .0003$).

\paragraph{Simulator Sickness.}
A significant overall effect was observed ($\chi^2(2) = 10.29$, $p = .006$), but post-hoc comparisons did not yield significant pairwise differences after correction.

\paragraph{System Usability (SUS).}
No significant differences were found across techniques ($\chi^2(2) = 2.53$, $p = .282$). All techniques received above-average SUS scores, suggesting generally acceptable levels of usability across conditions.

\paragraph{Final Preference.}
Most participants ($N=10$) ranked \textsc{Handows} highest in performance, 5 favored \textsc{Controller}, and none preferred \textsc{Virtual Hand}. 

\begin{table}[t]
\centering
\caption{Questionnaire results showing the mean (SD) for each input technique across all measured dimensions, including user experience (UEQ-S), workload (NASA-TLX), fatigue (Borg 6-20), simulator sickness, system usability (SUS) and final preference.}
\label{tab:questionnaire_results}
\resizebox{\columnwidth}{!}{%
\begin{tabular}{lccc}
\toprule
\textbf{Measure} & \textsc{Handows} & \textsc{Virtual Hand} & \textsc{Controller} \\
\midrule
\textbf{UEQ-S} \\
\quad Pragmatic Quality & 6.28 (0.75) & 4.25 (1.42) & 6.02 (0.69) \\
\quad Hedonic Quality   & 6.05 (0.67) & 4.57 (1.17) & 4.52 (1.46) \\
\quad Overall Score     & 6.04 (0.64) & 4.41 (1.16) & 5.40 (0.99) \\
\midrule
\textbf{NASA-TLX} \\
\quad Mental Demand     & 2.20 (0.77) & 3.40 (1.59) & 2.13 (1.46) \\
\quad Physical Demand   & 2.53 (1.19) & 5.00 (1.31) & 2.80 (1.66) \\
\quad Performance       & 1.80 (0.68) & 3.60 (1.64) & 2.00 (1.13) \\
\quad Effort            & 2.40 (1.16) & 4.80 (1.82) & 2.93 (1.80) \\
\quad Frustration       & 1.60 (0.63) & 3.80 (1.86) & 1.73 (0.88) \\
\midrule
\textbf{Borg 6-20} & 11.20 (1.73) & 15.20 (2.20) & 10.27 (2.36) \\
\midrule
\textbf{Simulator Sickness} & 3.67 (2.31) & 7.53 (4.94) & 3.13 (1.98) \\
\midrule
\textbf{System Usability (SUS)} & 73.50 (17.02) & 60.33 (15.31) & 67.83 (13.75) \\
%\midrule
%\textbf{Final Preference} & 66.7\% (10) & 0\% (0) & 33.3\% (5) \\
\bottomrule
\end{tabular}%
}
\end{table}

\subsection{Summary of Results}

\rev{The results across both objective and subjective measures consistently indicate that Handows offers substantial advantages in immersive window management. In particular, it significantly outperformed both Controller and Virtual Hand in window closure (Task 2) and scaling (Task 4), demonstrating faster task completion, reduced physical effort, and greater precision in scaling accuracy. For window selection (Task 1), Handows showed comparable performance to Controller and was notably faster than Virtual Hand. In contrast, for window positioning (Task 3), Controller achieved the best performance, while Handows did not show a statistically significant difference from Virtual Hand.}

\begin{figure*}[t]
\centering
\includegraphics[width=\linewidth]{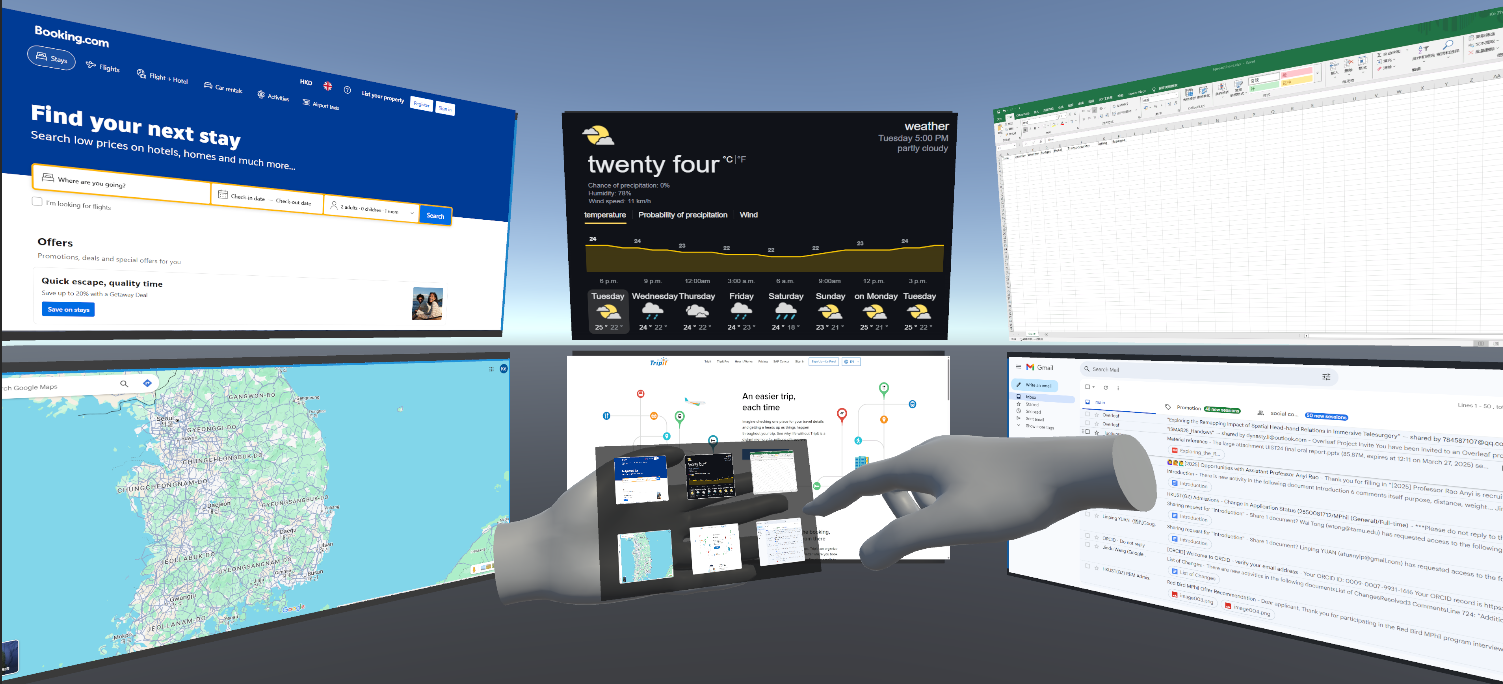}
\caption{Case study: Trip planning and budgeting using Handows, showcasing multitasking and dynamic window management in VR.}
\label{fig:casestudy}
\end{figure*}

\rev{The performance gap in Task 3 can be attributed to the nature of positioning tasks, which require continuous control and benefit from the precision and stability of dedicated ray-casting hardware. While the Controller condition offered reliable selection and tracking through button-based input, both Handows and Virtual Hand depended on hand gesture recognition, which may introduce latency and variability in tracking. Despite these limitations, Handows still achieved competitive performance without relying on external hardware—leveraging only passive haptics and familiar gesture paradigms. Moreover, the baseline conditions allowed unrestricted interaction with the full window surface, whereas Handows was intentionally designed around a compact, ergonomic palm interface. These findings suggest that while Handows does not surpass the Controller in every metric, it delivers robust and efficient interaction in most tasks, thus partially supporting our \textbf{H1}.}

\rev{In terms of ergonomic efficiency, Handows consistently resulted in reduced head rotation and hand movement across all tasks. Subjective measures further reinforce its benefits: participants reported lower mental and physical workload, less fatigue, and greater overall satisfaction with Handows compared to both baselines, especially Virtual Hand. Although the SUS scores were similar across techniques, Handows was rated highest in both pragmatic and hedonic quality dimensions, and was preferred by a majority of participants. Taken together, these findings provide strong evidence in support of \textbf{H2} and \textbf{H3}.}

\section{Case Study: Realistic Multitasking Scenario}

To explore the applicability of Handows, we conducted a case study involving a realistic multitasking scenario. The goal was to examine how users employed Handows to perform continuous window management operations in the context of a structured, goal-driven workflow, simulating common productivity scenarios in VR.

\subsection{Scenario and Method}

We developed a \emph{Trip Planning and Budgeting} scenario in which participants gathered, organized, and synthesized information across multiple spatial browser windows (see Figure~\ref{fig:casestudy}). The task involved sequential steps such as reviewing emails, checking weather forecasts, booking flights and accommodations, exploring restaurants and attractions, and compiling the final itinerary. This scenario was designed to elicit continuous use of all four core window operations (selection, closure, positioning, and scaling), which were previously evaluated in our user study, thus allowing us to assess the real-world fluidity and adaptability of Handows in a dynamic VR workflow.

8 right-handed participants (5 male, 3 female), aged 20--41 ($M=24.75$, $SD=7.63$) were recruited in this case study. Their familiarity with VR varied, with an average rating of ($M=2.75$, $SD=1.75$) on a 7-point Likert scale (1 = very unfamiliar, 7 = very familiar). None had taken part in the previous user study. 

The case study was conducted using the same apparatus as in the earlier tasks. After a brief training session with a blank layout to practice the core operations, participants completed the trip planning scenario using only \emph{Handows}, which took approximately 10 minutes. Post-task feedback on ease of use and enjoyment was collected via two 7-point Likert-scale questionnaires (1 = strongly disagree, 7 = strongly agree), as well as semi-structured interviews. The interviews focused on workflow adaptability, visual attention patterns, and overall user experience. 

\subsection{Results}

We analyzed the case study data using an inductive thematic analysis~\cite{braun2006using}, which revealed three major themes: \textit{(1) Workflow Adaptability and Efficiency}, \textit{(2) Visual Attention Strategies}, and \textit{(3) Perceived Usability and Suggestions for Improvement}. We also report quantitative ratings on ease of use and enjoyment to complement the qualitative findings.

\subsubsection{Theme 1: Workflow Adaptability and Efficiency}

All participants ($N=8$) agreed that \textit{Handows} fit naturally into the multitasking scenario. They described the system as ``intuitive,'' ``convenient,'' and easy to integrate into continuous window operations. Several participants noted that the design resembled familiar mobile interactions. For example, P1 shared: ``It feels like managing windows on my phone, and it aligns with my operating habits,'' while P7 commented: ``The hand panel simplifies the process and fits my intuitive operations.'' Participants also highlighted that the unified access to multiple operations enhanced efficiency. P3 mentioned: ``Multi-operations reduce the time spent switching between tools, making interactions more efficient.'' Others appreciated the reduction in spatial workload. As P4 explained, ``It reduces the fatigue of large-scale spatial movements, and the operation is easy to understand and use.'' P5 echoed this sentiment: ``The system covers all the window operations I wanted to perform in VR while reducing my movement range.''

When comparing \textit{Handows} to prior experience with VR controllers, three participants emphasized its flexibility and accessibility. For example, P6 noted: ``The hand panel follows the user's position, allowing manipulation from any angle, which makes it more fluid and accessible.''

\subsubsection{Theme 2: Visual Attention Strategies}

Participants reported a consistent gaze pattern during early interaction: they looked up to locate a window, looked down to execute the command using \textit{Handows}, and finally looked up again to verify the result. This sequence was especially prominent in selection, closure, and positioning tasks. For the scaling operation, participants described a cyclical gaze behavior. They repeatedly looked up to estimate the desired size and down at the hand panel to adjust the scaling gesture. P6 explained, ``I still have to look at the panel when scaling, especially to confirm my finger posture at the start.''

With increased familiarity, all participants ($N=8$) were able to execute simple operations like selection and closure without referring to the hand panel, indicating a low visual-cognitive load. However, more complex tasks such as positioning and scaling continued to require visual confirmation. P1, P3, P5, and P7 remarked that positioning gestures were difficult to execute blindly, as they required precise start and end finger placement. Similarly, P1, P5, and P6 emphasized the importance of finger alignment when performing the multi-finger scaling gesture.

\subsubsection{Theme 3: Perceived Usability and Suggestions for Improvement}

Participants generally found \textit{Handows} easy to use and enjoyable, rating its \textit{ease of use} at $(M = 5.25,\ SD = 0.46)$ and \textit{enjoyment} at $(M = 6.63,\ SD = 0.52)$ on a 7-point Likert scale. The mobile-inspired interaction was frequently described as ``intuitive'' and ``familiar'' (P1: ``It feels like managing windows on my phone.''; P7: ``The panel fits my intuitive operations well.''). While basic operations were smooth, more complex gestures posed challenges. Scaling was noted as physically tiring due to palm instability and recognition issues, and swapping felt ``less smooth, especially when multiple windows were involved'' (P5). Some participants also wished for features like reopening closed windows (P3, P5) or more layout customization (P1, P4).

Suggested improvements included scaling speed adjustment (P7), visual feedback for positioning (P4), and recovery options for closed windows (P8). These comments reflect a desire to maintain Handows' simplicity while enhancing control and flexibility in complex scenarios.

\section{Discussion}

Our study demonstrates that \textit{Handows} supports effective and efficient window management in VR, not only across isolated tasks but also within realistic multitasking workflows. By combining embodied input with familiar interaction metaphors, Handows bridges the gap between physical ergonomics and digital flexibility. In this section, we synthesize findings from both controlled experiments and the case study, discussing how \textit{Handows} enables task integration, supports embodied interaction, and reveals opportunities for future refinement.

\subsection{Task Integration and Multitasking Support}

One of the key contributions of \textit{Handows} lies in its capacity to facilitate fluid task switching across multiple window operations. In the four-task study, participants consistently performed selection and closure more quickly and with less head and hand movement compared to mid-air and controller-based alternatives. Tasks involving positioning and scaling, although more demanding, achieved high precision and benefited from the physical anchoring of the palm interface. Notably, scaling deviation was lowest in \textit{Handows} ($1.39\%$), supporting its suitability for fine-grained control.

The case study further validated \textit{Handows} in an open-ended multitasking scenario. Participants described the system as ``intuitive'' and ``mobile-like,'' often drawing parallels with familiar smartphone workflows (P1, P7). They organically developed spatial strategies (e.g., prioritizing central windows, offloading peripheral content, and sequencing closure and scaling) to maintain clarity and reduce cognitive clutter. These behaviors suggest that \textit{Handows} not only supports discrete interactions but also enables higher-level workflow structuring. Even in the absence of explicit instruction, users adapted their own heuristics for spatial layout and operational sequencing, reinforcing the system's learnability and flexibility.

\subsection{Embodied Design and Familiarity as Enablers}

The performance and subjective results across both studies point to three key design considerations underlying \textit{Handows}' success: body-centric interaction, spatial miniaturization, and interaction familiarity. First, anchoring window control on the non-dominant palm provides proprioceptive stability and passive haptic cues. Users rated the impact of passive haptic feedback from touching the hand during \textit{Handows} operation on operational stability as high, with an average score of 5.93 out of 7 ($SD = 0.88$). This embodiment reduced reliance on external spatial reference points, allowing for more compact movements and lower head rotation across tasks. Participants in the case study also noted that ``the hand panel follows the user's position,'' making it easier to operate from varying viewpoints (P6).

Second, the spatial miniaturization of interaction surfaces helped consolidate functions into a single accessible region. This design choice eliminated the need for large mid-air motions, improving speed and reducing fatigue. While scaling and positioning required visual confirmation, selection and closure were often performed without visual dependence after minimal training.

Third, the use of well-established mobile gestures (e.g., taps, swipes, pinches) allowed participants to transfer prior experience from 2D devices into a 3D context. This was supported by subjective ratings assessing the transferability of experience from traditional computing devices, where a Friedman test revealed significant differences ($\chi^2(2) = 20.63$, $p < .0001$). Handows scored highest at 6.13 out of 7 ($SD = 0.74$), significantly surpassing Controller at 4.13 ($SD = 1.73$, $p = 0.0005$) and Virtual Hand at 2.80 ($SD = 1.15$, $p < 0.0001$). This reuse of motor schemas was evident in both studies and aligns with the high ratings of \textit{Ease of Use} and \textit{enjoyment} in the case study. As P7 summarized, ``the [\textit{palm-based}] panel fits my intuitive operations well,'' highlighting how gesture familiarity directly enhanced usability.

\subsection{Limitations and Future Work}

Despite these advantages, our findings also surface areas for improvement. Participants reported occasional inaccuracies with high-precision gestures, particularly scaling. We attribute these challenges to a combination of user-perceived palm instability and the inherent limitations of current hardwares and hand tracking, which can struggle with the stable recognition of fine-grained movements and maintaining tracking at the periphery of camera views. \rev{We anticipate that newer headsets like the Meta Quest 3, equipped with depth sensors and passthrough tracking, will enhance tracking robustness and extend the effective interaction space, enabling more fluid and reliable window management. Beyond tracking, adapting the current Handows prototype into a full-featured window management system will require dynamic layout adjustments to accommodate increasing numbers of windows and offer greater configurability. }Participants also suggested specific enhancements to support complex workflows, such as options to reopen closed windows (P3, P5), improved layout customization (P1, P4), and recovery mechanisms for handling interaction errors (P8).

\rev{Future studies could incorporate individual hand measurements to investigate how physical differences affect interaction comfort and stability, thereby informing more personalized interface designs. Introducing the movement patterns of the non-dominant hand would offer a deeper understanding of user motor behavior and the ergonomic implications of window management interaction. Our case study served as an initial simulation of continuous window operations in a realistic scenario; future work may extend this by integrating interactions with window content and combining Handows with complementary input modalities. This would enable systematic future studies, including formal evaluations of gesture input accuracy and other quantitative measurements, enhancing both robustness and usability validation.}

Beyond these system-level improvements, another promising direction lies in understanding how user interaction patterns evolve with increased familiarity. While selection and closure were often performed without hand-panel observation, positioning and scaling remained visually anchored. Longitudinal studies could examine whether users develop at-a-glance strategies that rely more on spatial muscle memory and less on visual confirmation, thus potentially unlocking faster and more immersive workflows. Participants also proposed improvements to feedback and flow, for example, visual animations during positioning (P4) and dynamic scaling rate control (P7), which could support smoother transitions and reduce operational friction. Finally, integrating multimodal input (e.g., eye tracking for target locking or voice commands for window grouping) may further expand the bandwidth of palm-based interaction~\cite{lu_1,davari_1,pfeuffer_2}.

Finally, while our study focused on VR, the implementation of \textit{Handows} should also generalize well to MR settings. Its compact form factor, embodied layout, and reliance on familiar gestures make it especially suitable for productivity-focused MR environments~\cite{gao2023vr}. Future work should explore collaborative extensions, adaptive palm interfaces based on user context, and integration with virtual workspace platforms.

\section{Conclusion}

In this paper, we have presented \textit{Handows}, a palm-based window management system for virtual reality that integrates miniature spatial interfaces, body-centric interaction, and familiar gesture paradigms. Designed to reduce physical strain and improve workflow efficiency, Handows supports core window operations (i.e., selection, closure, positioning, and scaling) within the user's proprioceptive space using smartphone-inspired gestures.

Through a user study ($N$=15), we demonstrated that Handows outperforms common VR interaction techniques in task efficiency, precision, and user satisfaction, while significantly reducing physical effort. A complementary case study further validated its adaptability in realistic multitasking scenarios, where users employed strategic layout behaviors and reported high engagement. These findings underscore the value of embedding spatial interfaces into the body for fluid, low-effort VR interaction.

In the future, we see potential for extending Handows with adaptive layouts, multimodal input, and support for more complex workflows. More broadly, this work highlights how transplanting familiar interaction models onto embodied surfaces can support the design of VR systems that are both powerful and accessible.

%% if specified like this the section will be omitted in review mode
\acknowledgments{%
Xiang Li is supported by the China Scholarship Council (CSC) International Cambridge Scholarship (No. 202208320092). We thank all anonymous reviewers for their valuable feedback and our participants for their time and contributions.
}

\bibliographystyle{abbrv_doi_hyperref}

\bibliography{template}

% \appendix % You can use the `hideappendix` class option to skip everything after \appendix

% \section{About Appendices}
% Refer to \cref{sec:appendices_inst} for instructions regarding appendices.

% \section{Troubleshooting}
% \label{appendix:troubleshooting}

% \subsection{ifpdf error}

% If you receive compilation errors along the lines of \texttt{Package ifpdf Error: Name clash, \textbackslash ifpdf is already defined} then please add a new line \verb|\let\ifpdf\relax| right after the \verb|\documentclass[journal]{vgtc}| call.
% Note that your error is due to packages you use that define \verb|\ifpdf| which is obsolete (the result is that \verb|\ifpdf| is defined twice); these packages should be changed to use \verb|ifpdf| package instead.

% \subsection{\texttt{pdfendlink} error}

% Occasionally (for some \LaTeX\ distributions) this hyper-linked bib\TeX\ style may lead to \textbf{compilation errors} (\texttt{pdfendlink ended up in different nesting level ...}) if a reference entry is broken across two pages (due to a bug in \verb|hyperref|).
% In this case, make sure you have the latest version of the \verb|hyperref| package (i.e.\ update your \LaTeX\ installation/packages) or, alternatively, revert back to \verb|\bibliographystyle{abbrv-doi}| (at the expense of removing hyperlinks from the bibliography) and try \verb|\bibliographystyle{abbrv-doi-hyperref}| again after some more editing.

\end{document}